\documentclass[12pt,epsf]{article}
\usepackage{graphicx,amsmath,amssymb}
\begin{document}

\def\a{\alpha}
\def\b{\beta}
\def\c{\varepsilon}
\def\d{\delta}
\def\e{\epsilon}
\def\f{\phi}
\def\g{\gamma}
\def\h{\theta}
\def\k{\kappa}
\def\l{\lambda}
\def\m{\mu}
\def\n{\nu}
\def\p{\psi}
\def\q{\partial}
\def\r{\rho}
\def\s{\sigma}
\def\t{\tau}
\def\u{\upsilon}
\def\v{\varphi}
\def\w{\omega}
\def\x{\xi}
\def\y{\eta}
\def\z{\zeta}
\def\D{\Delta}
\def\G{\Gamma}
\def\H{\Theta}
\def\L{\Lambda}
\def\F{\Phi}
\def\P{\Psi}
\def\S{\Sigma}

\def\o{\over}
\def\beq{\begin{eqnarray}}
\def\eeq{\end{eqnarray}}
\newcommand{\gsim}{ \mathop{}_{\textstyle \sim}^{\textstyle >} }
\newcommand{\lsim}{ \mathop{}_{\textstyle \sim}^{\textstyle <} }
\newcommand{\vev}[1]{ \left\langle {#1} \right\rangle }
\newcommand{\bra}[1]{ \langle {#1} | }
\newcommand{\ket}[1]{ | {#1} \rangle }
\newcommand{\EV}{ {\rm eV} }
\newcommand{\KEV}{ {\rm keV} }
\newcommand{\MEV}{ {\rm MeV} }
\newcommand{\GEV}{ {\rm GeV} }
\newcommand{\TEV}{ {\rm TeV} }
\def\diag{\mathop{\rm diag}\nolimits}
\def\Spin{\mathop{\rm Spin}}
\def\SO{\mathop{\rm SO}}
\def\O{\mathop{\rm O}}
\def\SU{\mathop{\rm SU}}
\def\U{\mathop{\rm U}}
\def\Sp{\mathop{\rm Sp}}
\def\SL{\mathop{\rm SL}}
\def\tr{\mathop{\rm tr}}

\def\IJMP{Int.~J.~Mod.~Phys. }
\def\MPL{Mod.~Phys.~Lett. }
\def\NP{Nucl.~Phys. }
\def\PL{Phys.~Lett. }
\def\PR{Phys.~Rev. }
\def\PRL{Phys.~Rev.~Lett. }
\def\PTP{Prog.~Theor.~Phys. }
\def\ZP{Z.~Phys. }

\newcommand{\bea}{\begin{eqnarray}}   
\newcommand{\eea}{\end{eqnarray}}
\newcommand{\bear}{\begin{array}}  
\newcommand {\eear}{\end{array}}
\newcommand{\bef}{\begin{figure}}  
\newcommand {\eef}{\end{figure}}
\newcommand{\bec}{\begin{center}}  
\newcommand {\eec}{\end{center}}
\newcommand{\non}{\nonumber}  
\newcommand {\eqn}[1]{\beq {#1}\eeq}
\newcommand{\la}{\left\langle}  
\newcommand{\ra}{\right\rangle}
\newcommand{\ds}{\displaystyle}
\def\SEC#1{Sec.~\ref{#1}}
\def\FIG#1{Fig.~\ref{#1}}
\def\EQ#1{Eq.~(\ref{#1})}
\def\EQS#1{Eqs.~(\ref{#1})}
\def\GEV#1{10^{#1}{\rm\,GeV}}
\def\MEV#1{10^{#1}{\rm\,MeV}}
\def\KEV#1{10^{#1}{\rm\,keV}}
\def\lrf#1#2{ \left(\frac{#1}{#2}\right)}
\def\lrfp#1#2#3{ \left(\frac{#1}{#2} \right)^{#3}}


\baselineskip 0.7cm

\begin{titlepage}

\begin{flushright}
UT-12-35\\
TU-920\\
IPMU12-0171\\
\end{flushright}

\vskip 1.35cm
\begin{center}
{\large \bf 
	Eluding the Gravitino Overproduction in Inflaton Decay
}
\vskip 1.2cm
Kazunori Nakayama$^{a,c}$,
Fuminobu Takahashi$^{b,c}$
and 
Tsutomu T. Yanagida$^{c}$

\vskip 0.4cm

{\it $^a$Department of Physics, University of Tokyo, Tokyo 113-0033, Japan}\\
{\it $^b$Department of Physics, Tohoku University, Sendai 980-8578, Japan}\\
{\it $^c$Kavli Institute for the Physics and Mathematics of the Universe,
University of Tokyo, Kashiwa 277-8568, Japan}

\vskip 1.5cm

\abstract{
It is known that gravitinos are non-thermally produced in inflaton decay processes, which excludes many inflation models for
a wide range of the gravitino mass.  We find that the constraints from the gravitino overproduction 
can be greatly relaxed if the supersymmetry breaking field is much lighter than the inflaton, and if the dynamical scale of the supersymmetry breaking
is higher than the inflaton mass. In particular, we show that many inflation models then become consistent with the pure gravity mediation 
with $O(100)$\,TeV gravitino which naturally explains the recently observed Higgs boson mass of about 125\,GeV.
}
\end{center}
\end{titlepage}

\setcounter{page}{2}

\section{Introduction}

Recent discovery of the standard-model-like Higgs boson particle with mass about 125\,GeV at the LHC~\cite{Higgs}
may indicate relatively high-scale supersymmetry (SUSY) 
where the SUSY particle masses are of  order  $100$\,TeV~\cite{Okada:1990gg,Giudice:2011cg}.
In particular, the observed Higgs boson mass can be naturally explained in 
the so-called pure gravity mediation model~\cite{Ibe:2011aa}, where
sfermion masses as well as the gravitino mass are $O(100)$\,TeV, whereas gaugino masses are  $O(100)$\,GeV
generated by the anomaly-mediated SUSY breaking (AMSB) effect~\cite{Giudice:1998xp}. 
In the most parameter space,  the lightest SUSY particle (LSP) is the Wino.
Although the thermal relic density of the Wino is too small to account for the observed dark matter (DM) 
for the Wino lighter than $\sim 2.7$\,TeV~\cite{Hisano:2006nn}, it is also produced by the decay of the gravitino.
Since the gravitino is heavy enough to decay before big-bang nucleosynthesis (BBN), it does not spoil the success of BBN.
If the reheating temperature after inflation, $T_{\rm R}$, is around $10^9$--$10^{10}$\,GeV,
the non-thermal Wino  can explain the present DM abundance.
Such high reheating temperature is also consistent with thermal leptogenesis scenario~\cite{Fukugita:1986hr}.

While this is an attractive scenario, it is not trivial whether it  is consistent with known inflation models.
In a series of works~\cite{Endo:2006zj,Kawasaki:2006gs,Asaka:2006bv,Dine:2006ii,Endo:2006tf,
Endo:2006qk,Endo:2007ih,Endo:2007cu,Endo:2007sz},
it was revealed that the inflaton generally decays into gravitinos and these non-thermally produced gravitinos severely constrain
inflation models.
Even if the gravitino is as heavy as $O(100)$\,TeV, too many gravitinos would result in the LSP overproduction, which 
severely restricts inflation models.
The other aspects of the high-scale SUSY breaking in the context of inflation models is that the inflaton dynamics may be spoiled
or significantly modified by the existence of the constant term in the superpotential~\cite{Buchmuller:2000zm,Senoguz:2004vu,Nakayama:2010xf} or by the radiative correction to the inflaton potential~\cite{Nakayama:2011ri}.

One way to suppress the gravitino production in inflaton decay is to assign some charge to the SUSY breaking field $z$.
Then some of the dangerous terms in the K\"ahler potential, $K\sim |\phi|^2 z,~|\phi|^2 zz$, where $\phi$ denotes the inflaton, 
can be forbidden. Those operators are indeed suppressed in the low energy if $m_z \gg m_{3/2}$,
because the vacuum expectation value (VEV) of $z$ is then negligibly small. 
This is easily achieved in the dynamical SUSY breaking scenario. Interestingly,
gaugino masses are successfully generated by the AMSB contribution in the pure gravity mediation model,  
even  if $z$ is charged under a certain symmetry. In fact, since the F-term of $z$ develops VEV
there is still a mixing between $\phi$ and $z$, which induces the inflaton decay into the gravitinos. 
The rate, however, is significantly suppressed if $m_z \ll m_\phi$, where $m_z$ and $m_\phi$ denote
the mass of $z$ and $\phi$, respectively~\cite{Dine:2006ii,Endo:2006tf}. 

The problem is that if  the inflaton mass is larger than the dynamical SUSY breaking scale $\Lambda$, 
it can decay into hadrons in the hidden sector, which eventually produce  many gravitinos~\cite{Endo:2006qk,Endo:2007ih}.
Thus, a guess is that the gravitino production is suppressed if the following relation is satisfied :
\beq
m_{3/2} \ll m_z \ll m_\phi \lesssim \Lambda.
\label{relation}
\eeq
This requires a hierarchy between $m_z$ and $\Lambda$, which can be easily realized in some dynamical SUSY breaking scenarios,
as we shall see later.
Interestingly enough, the SUSY breaking scale $\Lambda \sim \sqrt{m_{3/2}M_P}$ is close to the inflaton mass in many inflation models
for $m_{3/2}\sim O(100)$\,TeV.
Thus we have much chance to suppress the gravitino overproduction in the high-scale SUSY scenario.

We note however that, if the mass of $z$ is too light, the gravitino production from the coherent oscillations of $z$ becomes
non-negligible. Therefore, it is important to take into account all these contributions to see to what extent the constraints
on the inflation models can be relaxed.


Lastly let us clarify the difference of the present paper from Ref.~\cite{Endo:2007cu}.
In Ref.~\cite{Endo:2007cu}, the relation (\ref{relation}) was assumed to avoid the gravitino 
overproduction in the gravity and gauge mediation, and the allowed region for the single-field new inflation was studied. 
In the present work, we shall derive the constraints on the general inflation 
model parameters for the case of heavy gravitino.

This paper is organized as follows.
In Sec.~\ref{sec:grav} we summarize the inflaton decay rate into the gravitino and the resulting gravitino abundance.
In Sec.~\ref{sec:Polonyi}, we discuss the Polonyi problem in dynamical SUSY breaking models
and show that the gravitino production can be indeed suppressed in an explicit SUSY breaking model.
We conclude in Sec.~\ref{sec:conc}.

\section{Non-thermal gravitino production from inflaton decay}  \label{sec:grav}

We assume dynamical SUSY breaking where SUSY is spontaneously broken by the strong dynamics 
at the scale $\Lambda$. A concrete model will be given later. 
Discussion in this section does not depend on details of the  dynamical SUSY breaking models.
Below the scale $\Lambda$, the SUSY breaking field $z$ has a superpotential of the form 
\beq
W = \mu^2 z   + W_0,
\eea
where $\mu$ represents the SUSY breaking scale, and the constant $W_0 \simeq m_{3/2} M_P^2$ 
is fixed so that the cosmological constant almost vanishes. The F-term of $z$ is given by
$F_z \simeq - \mu^2 \simeq \sqrt{3} m_{3/2} M_P$, and SUSY is indeed broken.
The $z$  obtains a non-SUSY mass through the following non-renormalizable operator in the K\"ahler potential,
\begin{equation}
	K \;\supset\; -\frac{|z|^4}{\tilde\Lambda^2}.   \label{Kz4}
\end{equation}
Here $\tilde\Lambda$ is some cutoff scale, which is roughly equal to $\Lambda$ if $z$ itself is involved in the
strong dynamics, while it can be much larger than $\Lambda$ if $z$ is weakly coupled to the strong sector as shown explicitly in Sec.~\ref{sec:DSB}.
It generates the mass of $z$ as $m_z^2 = 4|F_z|^2/\tilde\Lambda^2$. We assume $m_z \gg m_{3/2}$ so that the VEV of $z$
is suppressed by $m_{3/2}^2/m_z^2$. Hereafter we assume that $z$ is charged under some symmetry, such as global U(1), which is spontaneously broken by
the strong dynamics in the hidden sector.

Let us consider the mixing of inflaton, which is denoted by $X$ or $\phi$ in the following, 
and SUSY breaking field $z$.
As an example, we consider the following K\"ahler and super-potentials:
\bea
K &=& |\phi|^2+|X|^2 +|z|^2  -\frac{|z|^4}{\tilde\Lambda^2},\\
W &=& X(g\phi^n - v^2) + \mu^2 z  + W_0,
	\label{spp}
\eea
where the first term in $W$ corresponds to the inflaton sector with $g$ being the coupling constant and $v$ the constant
giving the inflation energy scale. At the potential minimum, $\phi$ develops a VEV, $\la \phi \ra \equiv |v^2/g|^{1/n}$,
while $X$ sits near the origin. Note that $\phi^n$ can be replaced with $(\phi\bar\phi)^{n/2}$, but the following discussion does not change due to this choice. 
This class of inflation models includes the hybrid $(n=2)$~\cite{Copeland:1994vg} and smooth-hybrid inflation~\cite{Lazarides:1995vr} 
as well as the new inflation model $(n\geq 4)$~\cite{Izawa:1996dv,Asaka:1999jb}. Also,
the following arguments can be applied to the chaotic inflation model~\cite{Kawasaki:2000yn}
without a discrete symmetry on $X$ and $\phi$.

Around the potential minimum, $\phi$ and $X$ get maximally mixed with each other to form mass eigenstates, $\Phi_\pm
\equiv ( \phi \pm X^\dag)/\sqrt{2}$, in the presence of $W_0$~\cite{Kawasaki:2006gs}. The inflaton mass is (approximately)
given by $m_\phi = ng \langle\phi\rangle^{n-1}$.
This mixing is meaningful as long as the decay rates of $\phi$ and $X$ are smaller than $m_{3/2}$, which is assumed 
in the following.\footnote{Otherwise, too many gravitinos are thermally produced.}

From the supergravity scalar potential, we find the mixing of $X$ and $z$ as
\begin{equation}
	V = e^{K/M_P^2}\left[ K_{i\bar j}^{-1}(D_i W)(D_{\bar j}\bar W)  -3\frac{|W|^2}{M_P^2}\right] 
	\supset \frac{m_\phi\langle\phi\rangle \mu^2}{M_P^2}Xz^\dagger + {\rm h.c.}.
\end{equation}
The mixing angle between $X$ and $z$ is approximately given by
\begin{equation}
	\theta \;\simeq\; \left|\frac{m_\phi\langle\phi\rangle F_z}{M_P^2(m_\phi^2-m_z^2)}\right|
	\simeq 
	\begin{cases}
		\displaystyle\frac{\sqrt{3}m_{3/2}\langle\phi\rangle }{m_\phi M_P} & {\rm~for~}m_\phi \gg m_z,\\
		\displaystyle\frac{\sqrt{3}m_{3/2}m_\phi\langle\phi\rangle }{m_z^2 M_P} & {\rm~for~}m_\phi \ll m_z.
	\end{cases}
\end{equation}
Thus, the effective mixing angle between $\Phi_\pm$ and $z$ is given by $\theta/\sqrt{2}$.

The inflaton decay into the gravitino is induced by the operator (\ref{Kz4}).
It leads to the following term in the Lagrangian
\begin{equation}
	\mathcal L \supset -2\frac{F_z^\dagger}{\tilde\Lambda^2}z^\dagger \tilde z\tilde z +{\rm h.c.},
\end{equation}
where $\tilde z$ denotes the goldstino, which is eaten by the gravitino through the super Higgs mechanism.
This operator  induces the $z$ decay into the goldstino pair with the decay rate
\begin{equation}
	\Gamma(z\to \tilde z\tilde z) \;\simeq\; \frac{1}{96\pi}\frac{m_z^5}{m_{3/2}^2M_P^2}.  
	\label{StoGG}
\end{equation}
As far as the inflaton mass is much heavier than the gravitino, we can estimate the inflaton decay 
into  gravitinos in the goldstino picture thanks to the equivalence theorem.  The inflaton decays into
a pair of goldstinos via the mixing with $z$, and the rate is given by
\begin{equation}
	\Gamma(\Phi \to \tilde z\tilde z) \;\simeq\; \frac{1}{32\pi} \lrfp{\theta}{\sqrt{2}}{2}  \frac{m_z^4}{|F_z|^2}m_\phi =
	\begin{cases}
		\displaystyle
		\frac{1}{64\pi}\left( \frac{m_z}{m_\phi} \right)^4\left( \frac{\langle\phi\rangle}{M_P} \right)^2 \frac{m_\phi^3}{M_P^2}
		& {\rm~for~}m_\phi \gg m_z,\\
		\displaystyle
		\frac{1}{64\pi}\left( \frac{\langle\phi\rangle}{M_P} \right)^2 \frac{m_\phi^3}{M_P^2}
		& {\rm~for~}m_\phi \ll m_z,
	\end{cases}
	\label{PhiSS}
\end{equation}
where $\Phi$ collectively denotes the inflaton mass eigenstates $\Phi_\pm$.
Therefore, the decay rate is suppressed for $m_\phi \gg m_z$. The precise form of the decay rate
is given in Appendix. 

Note that $z$ has a charge and hence terms such as $K \supset |\phi|^2 z$ and $ |\phi|^2 zz$ are forbidden,
which would otherwise induce the gravitino oveproduction.
However, no symmetry forbids the following non-renormalizable interaction between the inflaton and $z$:
\begin{equation}
	K \;\supset\; -c\frac{|\phi|^2 |z|^2}{M_P^2},
\end{equation}
where $c$ is a constant of order unity. This induces the inflaton decay into the scalar component of 
the SUSY breaking field as
\begin{equation}
	\Gamma(\Phi \to zz^\dagger) = \frac{c^2}{32\pi}\left( \frac{m_z}{m_\phi} \right)^4
	\left( \frac{\langle\phi\rangle}{M_P} \right)^2 \frac{m_\phi^3}{M_P^2}
	\left( 1- \frac{4m_z^2}{m_\phi^2} \right)^{1/2}.
\end{equation}
Since $z$  predominantly decays into the gravitino pair,
this process yields gravitinos with the same order of those from (\ref{PhiSS}).
Note also that the operator like $K\sim (|\phi|^2/M_P^2)(|z|^4/\tilde\Lambda^2)$ gives comparable rate with that given above.
See Appendix for the details.

If the inflaton is heavier than the dynamical scale $\Lambda$, the inflaton decays into
hadrons in the hidden sector, which also poses severe constraints on inflation models.
The decay proceeds through both tree-level~\cite{Endo:2006qk} and one-loop level~\cite{Endo:2007ih}, 
but the tree-level process
depends on the details of the SUSY breaking models, while the decay via
anomalies is more robust.
Assuming that the hidden hadron masses are given by $\Lambda$, the decay rate at one-loop level
is given by~\cite{Endo:2007ih,Endo:2007sz}
\begin{equation}
	\Gamma(\Phi \to {\rm hadron}) =
	\begin{cases}
		\displaystyle
		\frac{N_g\alpha_h^2}{512\pi^3}(\mathcal T_G-\mathcal T_R)^2\left( \frac{\langle\phi\rangle}{M_P} \right)^2 \frac{m_\phi^3}{M_P^2}
		& {\rm~for~}m_\phi \gtrsim 2\Lambda,\\
		\displaystyle
		0
		& {\rm~for~}m_\phi \lesssim 2\Lambda,
	\end{cases}
\end{equation}
where $\mathcal T_G$ and $\mathcal T_R$ are Dynkin index of the adjoint representation and and matter fields in the representation $R$,
$\alpha_h$ is the fine structure constant of the hidden gauge group and $N_g$ the number of generators of the gauge group.
We have assumed the minimal coupling between the inflaton sector and the hidden sector in the K\"ahler potential. 
For simplicity, we take $N_g \alpha_h^2(\mathcal T_G-\mathcal T_R)^2 =1$ in the numerical calculation. 
If this decay mode is open, the gravitino overproduction problem is severe
since each hidden hadron jets finally produce gravitinos.
As a result, we obtain the following condition for significantly relaxing the gravitino overproduction problem :
\begin{equation}
	 m_{3/2} \ll m_z \ll m_\phi \lesssim \Lambda.
	 \label{inequality}
\end{equation}
Actually, this condition is easily satisfied in a dynamical SUSY breaking model explained in Sec.~\ref{sec:DSB} (see Eq.~(\ref{mS})).
However, one should note that too light $m_z$ may lead to the Polonyi problem as shown later.

The gravitino abundance, in terms of the number-to-entropy ratio, $Y_{3/2}\equiv n_{3/2}/s$, is given by
\begin{equation}
	Y_{3/2}^{(\phi)}=\frac{3T_{\rm R}}{4m_\phi}
	\frac{ 2\Gamma(\Phi\to \tilde z\tilde z)+4\Gamma(\Phi\to zz^\dagger)+2N_{3/2}\Gamma(\Phi\to{\rm hadron}) }{\Gamma_{\rm tot}},
	\label{Ygrav}
\end{equation}
where $\Gamma_{\rm tot}$ is the total decay rate of the inflaton and it is related to the reheating 
temperature $T_{\rm R}$ as $\Gamma_{\rm tot} \equiv (\pi^2 g_*/90)^{1/2} T_{\rm R}^2/M_P$,
and $N_{3/2}$ represents the averaged number  of gravitinos per hidden hadron jet. 
We will take $N_{3/2}=1$ for simplicity. 

\begin{figure}
\begin{center}
\includegraphics[scale=1.6]{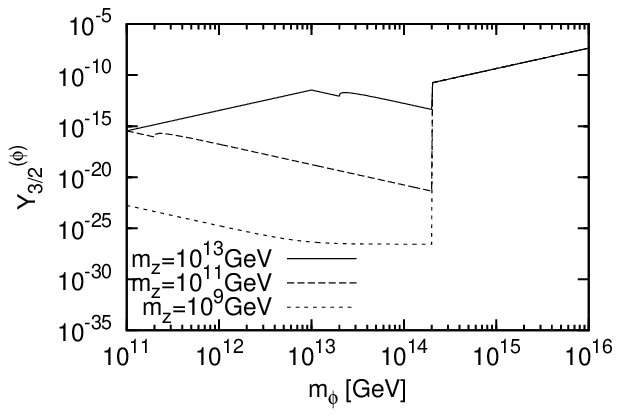}
\vskip 1cm
\includegraphics[scale=1.6]{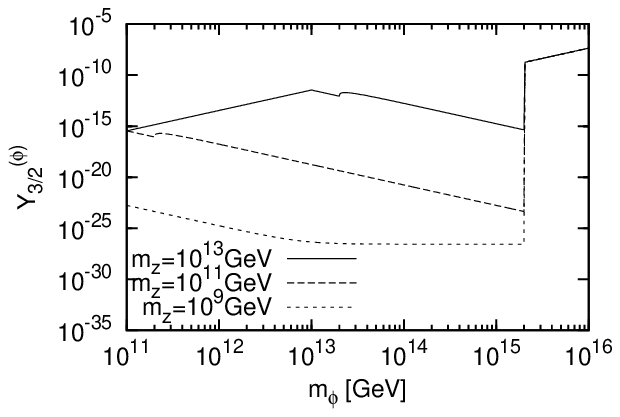}
\caption{ 
	Non-thermally produced gravitino abundance, $Y_{3/2}^{(\phi)}$, from inflaton decay as a function of inflaton mass $m_\phi$
	for several values of mass of the SUSY breaking field $m_z$.
	We have taken $\Lambda = 10^{14}$\,GeV (top panel) and $\Lambda = 10^{15}$\,GeV (bottom panel) for
	$\langle\phi\rangle=10^{15}$\,GeV and $T_{\rm R}=3\times 10^9$\,GeV.
	Note that $Y_{3/2}^{(\phi)}$ scales as $\propto T_{\rm R}^{-1}$ and $\propto \langle\phi\rangle^2$. The lines for $m_z = \GEV{9}$
	are flattened because of the kinetic mixing between $\phi$ and $z$. See Appendix for details. 
	}
\label{fig:Ygrav}
\end{center}
\end{figure}

Fig.~\ref{fig:Ygrav} shows non-thermally produced gravitino abundance, $Y_{3/2}^{(\phi)}$, 
from inflaton decay as a function of inflaton mass $m_\phi$ for several values of $m_z$.
We have taken $\Lambda = 10^{14}$\,GeV (top panel) and $\Lambda = 10^{15}$\,GeV (bottom panel) for
$\langle\phi\rangle=10^{15}$\,GeV and $T_{\rm R}=3\times 10^9$\,GeV.
It is clearly seen that the gravitino abundance is significantly reduced in the range $m_z \ll m_\phi < \Lambda$.
At large $m_\phi$, three lines coincide since the gravitino production is dominated by the inflaton decay into hidden hadrons.
One can read off the gravitino abundance for other values of $\langle\phi\rangle$ and $T_{\rm R}$
by noting that $Y_{3/2}^{(\phi)}$ simply scales as $\propto T_{\rm R}^{-1}$ and $\propto \langle\phi\rangle^2$.

\section{Constraints on inflation models in dynamical SUSY breaking}   \label{sec:Polonyi}

\subsection{Polonyi problem in dynamical SUSY breaking}

In this section we discuss the Polonyi problem in the dynamical SUSY breaking scenario.
Since the SUSY breaking field $z$ obtains a large mass and can have a charge,
the cosmological problem associated with the $z$ coherent oscillation is much weaker than the conventional Polonyi problem
in gravity-mediation models~\cite{Coughlan:1983ci}.
Still, however, there may be significant contributions to the gravitino abundance from the decay of the $z$ coherent oscillations.
Let us go into details.

Below the dynamical scale $\Lambda$, the potential of the Polonyi field $z$ can be written as\footnote{
In the hybrid inflation, there will be a linear term $\sim H_{\rm inf} \mu^2 \la X \ra_{\rm inf} z /M_P+ {\rm h.c.}$,
where $\la X \ra_{\rm inf}$ represents the inflaton field value during inflation. This however does not change
the argument. 
}
\begin{equation}
	V = bH^2|z|^2 + m_z^2|z|^2 - (2m_{3/2}\mu^2 z + {\rm h.c.}).
\end{equation}
where $H$ denotes the Hubble parameter and $b$ is a constant of order unity assumed to be positive.
Let us estimate the Polonyi abundance in the two cases : $H_{\rm inf} \gg m_z$ and $H_{\rm inf} \ll m_z$, where
$H_{\rm inf}$ denotes the Hubble scale during inflation.

First we discuss the case of $H_{\rm inf} \gg m_z$.
When $H$ is large enough, the minimum of $z$ is close to the origin.
It is expected that the $z$ begins to oscillate around the true minimum at $H\simeq m_z$ with an amplitude of
\begin{equation}
	\langle z\rangle = \frac{2\sqrt{3} m_{3/2}^2 M_P}{m_z^2}.  \label{SVEV}
\end{equation}
Thus the Polonyi abundance is given by
\begin{equation}
	\frac{\rho_z}{s} = 3T_{\rm R}\left(\frac{m_{3/2}}{m_z}\right)^4,
\end{equation}
where $T_{\rm R}$ is the reheating temperature and we have assumed $T_{\rm R} \lesssim \sqrt{m_z M_P}$.

Next we consider the opposite case, $H_{\rm inf} \ll m_z$.
In this case, $z$ already sits at the position close to the minimum during inflation.
The deviation from the true minimum at the end of inflation is estimated as
\begin{equation}
	|\delta z| \simeq \frac{2\sqrt{3} m_{3/2}^2 M_P}{m_z^2}\left( \frac{bH_{\rm inf}^2}{m_z^2} \right) .
\end{equation}
Since $m_\phi \gg m_z$, the Polonyi cannot track the change of the potential at the end of inflation and
oscillation of the Polonyi field is induced~\cite{Nakayama:2011wqa}.
Then the Polonyi abundance is given by\footnote{
On the other hand, if $m_\phi \ll m_z$, the change of the Polonyi potential is adiabatic with respective to its mass scale
and hence no significant oscillation is induced.
}
\begin{equation}
	\frac{\rho_z}{s} \simeq 3T_{\rm R}\left(\frac{m_{3/2}}{m_z}\right)^4 \left( \frac{b^2H_{\rm inf}^2}{m_z^2} \right).
\end{equation}

As shown in (\ref{StoGG}), the Polonyi dominantly decays into the gravitino pair.
The gravitino abundance produced by the Polonyi decay is calculated as
\begin{equation}
	Y_{3/2}^{(z)} = \frac{2}{m_z}\frac{\rho_z}{s} 
	\simeq 6\times 10^{-16} \epsilon \left( \frac{T_{\rm R}}{10^9\,{\rm GeV}} \right)
	\left( \frac{m_{3/2}}{100\,{\rm TeV}} \right)^4 \left( \frac{10^{9}\,{\rm GeV}}{m_z} \right)^5,
	\label{Ygrav_z}
\end{equation}
where
\begin{equation}
	\epsilon = \begin{cases}
		1 & {\rm for~~} H_{\rm inf} \gg m_z \\
		H_{\rm inf}^2/m_z^2  & {\rm for~~}H_{\rm inf} \ll m_z.
	\end{cases}
\end{equation}
Therefore, the contribution to the gravitino abundance from the $z$ coherent oscillations is
negligible for $m_z \gtrsim 10^{9}$\,GeV for $m_{3/2} \sim 10^{2}-10^3$\,TeV.
We assume this in the following.
Note that the VEV of $z$ (\ref{SVEV}) is smaller than $\Lambda$ in such a case,
hence the discussion so far remains valid. This should be contrasted to the analysis
of Ref.~\cite{Endo:2007cu}.

\subsection{A model of dynamical SUSY breaking}  \label{sec:DSB}

Here we give an example of dynamical SUSY breaking model : the IYIT model~\cite{Izawa:1996pk}
having a desired structure to suppress the gravitino overproduction.
We introduce chiral superfields $Q_i$ $(i=1-4)$, each of which transforms as a doublet representation under an SP(1) gauge group, 
which becomes strong at the dynamical scale $\Lambda$.
We also introduce six gauge singlets $z_{ij}$ ($z_{ij}=-z_{ji}$) which couples to $Q_i$ as follows :
\begin{equation}
	W = \lambda z_{ij} Q_i Q_j.   \label{WSQQ}
\end{equation}
This form of the coupling is ensured by SU(4)$_F$ flavor symmetry, under which both $Q_i$ and $z_{ij}$ are charged.
The strong dynamics enforces a constraint on the $QQ$ pair as ${\rm Pf}(Q_iQ_j) = \Lambda^4$.
This contradicts with the equation of motion of $z_{ij}$, $\partial W/\partial z_{ij}=0$. Hence, SUSY is broken dynamically.
As a result, one of the combination of $z_{ij}$, which we denote by $z$, obtains an $F$-term as
\begin{equation}
	F_z = \frac{\lambda \Lambda^2}{(4\pi)^2},
\end{equation}
where we have relied on the naive dimensional analysis~\cite{Luty:1997fk}.
Hereafter we assume that $z$ has a charge under some symmetry group.
For example, it can have a global U(1) symmetry under which $z$ and $QQ$ transform as
$z \to e^{i\theta} z$ and $(QQ) \to e^{-i\theta} (QQ)$.\footnote{
	This symmetry is anomalous under the gauge group and broken down to a discrete subgroup,
	which is spontaneously broken below the scale $\Lambda$.
	Hence there may be a domain wall problem. 
	This is avoided if the SUSY is already broken during inflation so that domain walls are inflated away, or if there are small
	explicit symmetry breaking terms that destabilize domain walls.
}
Since $F_z$ is related to the gravitino mass through the relation $F_z=\sqrt{3}m_{3/2}M_P$,
we can express the dynamical scale $\Lambda$ as
\begin{equation}
	\Lambda = 8\times 10^{12}\,{\rm GeV} \frac{1}{\sqrt{\lambda}}\left(\frac{m_{3/2}}{100\,{\rm TeV}}\right)^{1/2}.
	\label{Lambda}
\end{equation}
Notice that this is close to the inflaton mass scale for many inflation models.
The mass of $z$ is generated from the quantum corrected effective K\"ahler potential
\begin{equation}
	K \supset -\frac{\lambda^4}{16\pi^2}\frac{|z|^4}{\Lambda^2}.
	\label{KS4}
\end{equation}
Therefore, $\tilde\Lambda$ in (\ref{Kz4}) is related with $\Lambda$ through the relation 
$\tilde\Lambda=(4\pi/\lambda^2)\Lambda$.
This yields
\begin{equation}
	m_z = \frac{2\lambda^3}{(4\pi)^3}\Lambda.
	\label{mS}
\end{equation}
Thus $m_z$ is much smaller than the dynamical scale $\Lambda$ for $\lambda \ll 4\pi$,
while hadrons in hidden sector have masses of $\sim \Lambda$.
For fixed gravitino mass,  $\Lambda$ becomes larger and
$z$ becomes lighter as $\lambda$ decreases, and so, the gravitino production rate is suppressed (see Eq.~(\ref{PhiSS})).
This hierarchy between $m_z$ and $\Lambda$ has important implications on the
gravitino overproduction problem from inflaton decay.

Note that the superpotential (\ref{WSQQ}) induces the three-body inflaton decay into $z QQ$.
The decay rate is given by~\cite{Endo:2006qk}
\begin{equation}
	\frac{1}{3}\Gamma(\phi \to z QQ) =
	\frac{1}{2}\Gamma(\phi \to \tilde z\tilde QQ) =\Gamma(\phi \to z \tilde Q \tilde Q) = \frac{\lambda^2}{768\pi^3} 
	\left( \frac{\langle\phi\rangle}{M_P} \right)^2 \frac{m_\phi^3}{M_P^2},
\end{equation}
where $Q (\tilde Q)$ represents the scalar (fermionic) component.\footnote{
	Three body decays including the other heavier components of $z_{ij}$ are also possible for $m_\phi \gg \Lambda$.
	They will increase the gravitino abundance up to some numerical factor.
}
Gravitinos are produced by these processes and they should be added to the estimate (\ref{Ygrav}) as
\begin{equation}
	\delta Y_{3/2}^{(\phi)}=\frac{3T_{\rm R}}{4m_\phi}
	\frac{ (10+12N_{3/2})\Gamma(\phi\to z\tilde Q\tilde Q) }{\Gamma_{\rm tot}},
\end{equation}
for $m_\phi > 2\Lambda$.

\subsection{Constraint on inflation models}   \label{sec:const}

Now let us derive constraints on inflation models from the gravitino overproduction.
We consider the following SUSY inflation models :
new inflation~\cite{Izawa:1996dv,Asaka:1999jb,Nakayama:2011ri},
hybrid inflation~\cite{Copeland:1994vg,Nakayama:2010xf},
smooth-hybrid inflation~\cite{Lazarides:1995vr} and chaotic inflation~\cite{Kawasaki:2000yn}.
Since we are interested in the heavy gravitino scenario, gravitinos decay well before BBN.
The constraint comes from the requirement that LSPs produced by the decay of (non-)thermal gravitino
should not exceed the observed DM abundance : $m_{\rm LSP} (Y_{3/2}^{(\phi)} +Y_{3/2}^{\rm (th)}+Y_{\rm LSP}^{\rm (th)})< 4\times 10^{-10}$\,GeV,
where $Y_{3/2}^{\rm (th)}$ and $Y_{\rm LSP}^{\rm (th)}$ denote the abundance of thermal gravitinos and
the thermal relic abundance of the LSP, respectively~\cite{Bolz:2000fu} and $m_{\rm LSP}$ the LSP mass. 
Hereafter we assume the Wino LSP. Then, for the Wino mass lighter than $\sim 2.7$\,TeV, the thermal relic density
is too small to account for all the dark matter density.

\begin{figure}[]
\begin{center}
\includegraphics[scale=0.9]{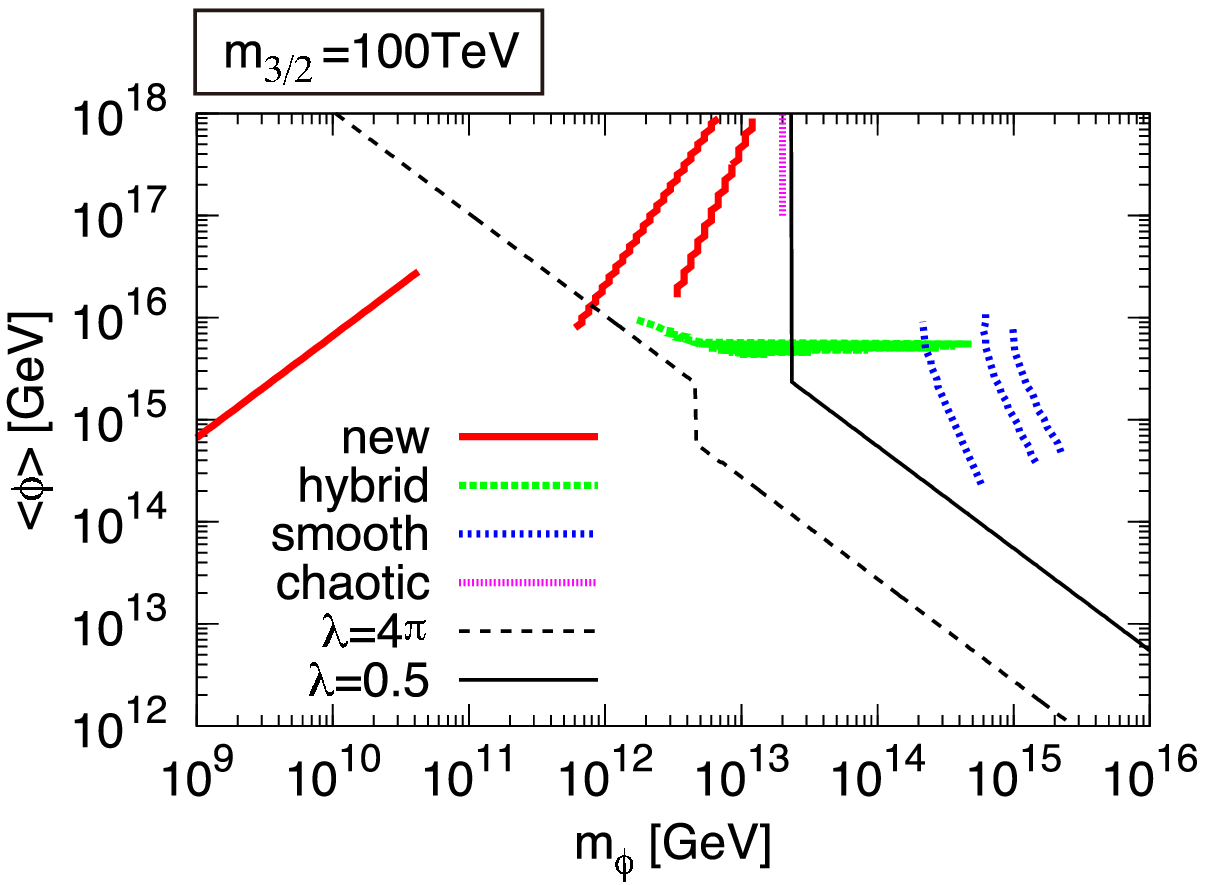}
\vskip 1cm
\includegraphics[scale=0.9]{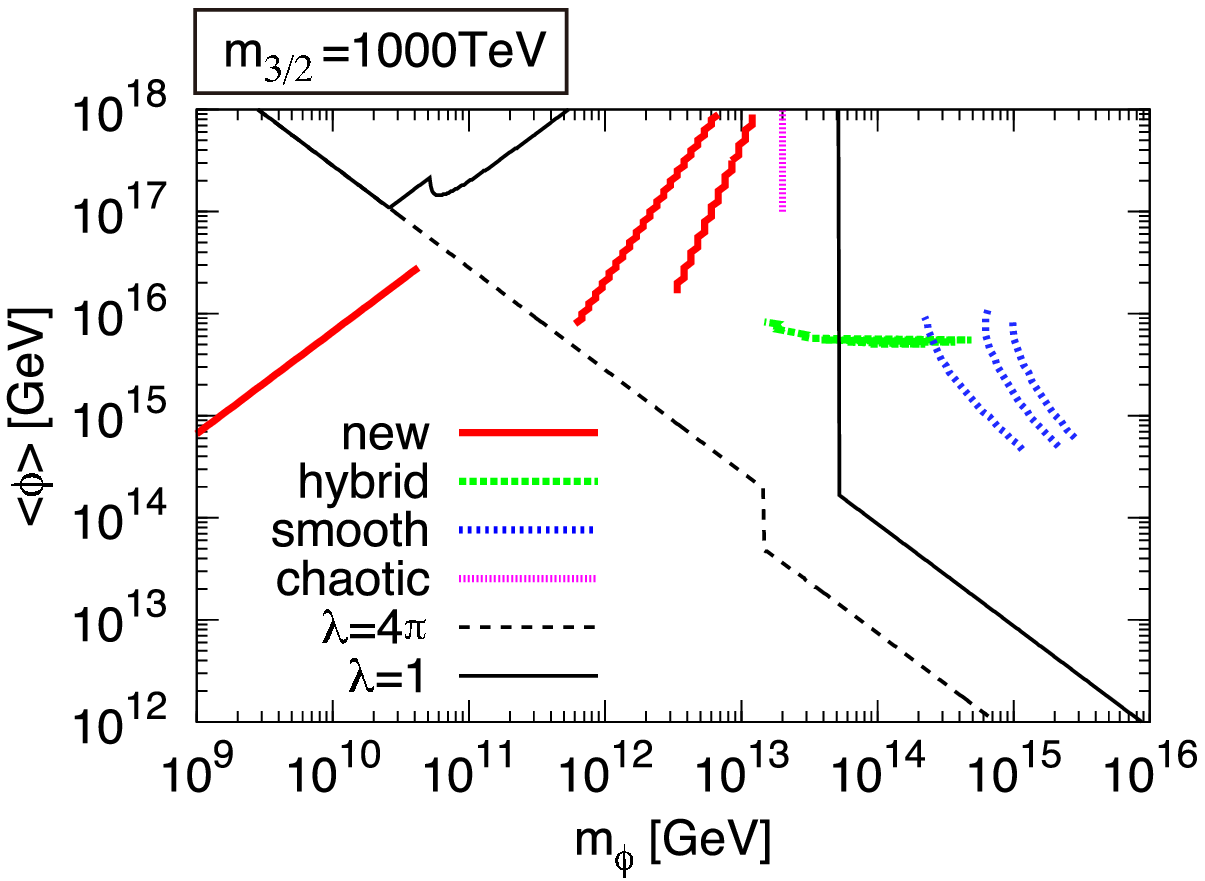}
\caption{ Constraint on inflation models on the $m_\phi$--$\langle\phi\rangle$ plane for several values of $\lambda$.
The region above the lines are excluded.
We have taken $m_{3/2}=100$\,TeV  and assumed the AMSB relation for the Wino mass ($\simeq 270$\,GeV)
in the top panel, while $m_{3/2}=10^3$\,TeV, and the Wino mass set to be $1$\,TeV in the bottom.
 We have fixed $T_{\rm R}$ so that the 
Winos produced by the decay of thermal gravitinos account for about half of the present DM abundance.
}
\label{fig:const_lam}
\end{center}
\end{figure}

Fig.~\ref{fig:const_lam} shows constraints on inflation models on $m_\phi$--$\langle\phi\rangle$ plane
for several values of $\lambda$ in the IYIT SUSY breaking model. We have taken $m_{3/2}=100$\,TeV  and 
assumed the AMSB relation for the Wino mass $m_{\tilde W} (\simeq 270\,{\rm GeV})$
in the top panel, while $m_{3/2}=10^3$\,TeV, and the Wino mass set to be $1$\,TeV in the bottom.
Note that the gaugino masses do not necessarily satisfy the AMSB relation in the pure gravity mediation~\cite{Ibe:2011aa}.
In particular, the Wino mass receives the Higgs-Higgsino loop contribution, and it can be a few times heavier (or
lighter) than the mass determined by the AMSB relation.
We have fixed $T_{\rm R}$ so that the Winos emitted by the decay of thermally produced gravitinos 
account for about half of the present DM abundance. 
The WMAP normalization~\cite{Komatsu:2010fb} on the density perturbation is satisfied for all inflation models.
We have included the effect of the constant term in the superpotential, $W_0$, on the inflaotn dynamics.
It changes the parameter space for the hybrid inflation model between $m_{3/2}=100$\,TeV and $10^3$\,TeV.
For the  new and smooth-hybrid inflation, three lines correspond to $n=4,6,8$ from left to right. 
It is seen that the constraint is significantly relaxed for small $\lambda$ since $m_z$ becomes small 
and the gravitino production rate gets suppressed by a factor of $\sim (m_z/m_\phi)^4$ for $m_z \ll m_\phi$.
It is remarkable that the hybrid inflation model and new inflation with $n>2$,
and even the chaotic inflation model without $Z_2$-symmetry may be allowed.

The abundance of the non-thermal gravitino is proportional to $m_{\tilde W} \la \phi \ra^2/T_{\rm R}$.
Thus, for the other parameters fixed, the constraints in the figure shift as $\sqrt{T_{\rm R}/m_{\tilde W}}$,
as long as $m_{\tilde W} (Y_{3/2}^{\rm (th)}+Y_{\tilde W}^{\rm (th)})$ do not exceed about half of the 
observed dark matter abundance.
For instance, if we decrease $T_{\rm R}$ by a factor of $10^2$, the constraint on $\la \phi \ra$
becomes severer by a factor of $10$ for the fixed inflaton mass. 
Note also that we cannot reduce the value of $\lambda$ further, since it tends to decrease the $z$ mass
and correspondingly the Polonyi-induced gravitino problem becomes severer (see Eq.~(\ref{Ygrav_z})).

\section{Conclusions}   \label{sec:conc}

We have revisited the issue of gravitino overproduction in inflaton decay in light of the 
recent discovery of the 125\,GeV Higgs boson, which implies relatively heavy gravitino : $m_{3/2}=10^2$--$10^3$\,TeV.
It is found that gravitino production rate is significantly suppressed in a dynamical SUSY breaking scenario,
if following conditions are met.
(1) The SUSY breaking field $z$ is charged under some symmetry, so that terms such as $|\phi|^2 z$ and $|\phi|^2 zz$ are forbidden.
(2) There is  hierarchy among the gravitino mass, the $z$ mass, $m_z$, and the dynamical scale $\Lambda$.
Then, the gravitino overproduction in inflation models with $m_{3/2} \ll m_z \ll m_\phi \lesssim \Lambda$ are 
greatly relaxed. 
Thus many inflation models are consistent with the SUSY breaking scenario with $m_{3/2}=10^2$--$10^3$\,TeV.
We have obtained the constraints on the inflation models in the pure gravity mediation assuming the IYIT SUSY breaking model.

\section*{Acknowledgments}

This work was supported by the Grant-in-Aid for Scientific Research on
Innovative Areas (No. 21111006  [KN and FT],  No.23104008 [FT], No.24111702 [FT]),
Scientific Research (A) (No. 22244030 [KN and FT], 21244033 [FT], 22244021 [TTY]), and JSPS Grant-in-Aid for
Young Scientists (B) (No.24740135) [FT].  This work was also
supported by World Premier International Center Initiative (WPI Program), MEXT, Japan.

\appendix
\section{Inflaton decay rate}

In this Appendix we summarize formulae for the inflaton decay rate into
a pair of the gravitinos and that into $z$. The inflaton $\phi$ is assumed to be stabilized
at $\phi = \la \phi \ra$ with a large SUSY mass, $m_\phi$. For simplicity we focus on a single-field
inflation. In the presence of $X$ as in \EQ{spp}, the mixing between the inflaton mass eigenstate(s)
with $z$ should be effectively multiplied with $1/\sqrt{2}$, because $\la \Phi_\pm \ra
 =\la  (\phi \pm X^\dag)/\sqrt{2} \ra \simeq \la \phi \ra /\sqrt{2}$. Therefore the decay rates in the text
are half of the followings. We adopt the Planck unit, unless the Planck scale
 is explicitly shown. 

\subsection{Decay into a pair of gravitinos}
%
%
%
We assume that  (\ref{inequality}) is satisfied, and that the $z$ is 
charged under some symmetry so that its VEV is suppressed by $m_{3/2}^2/m_z^2$.
Then, the decay rate of the inflaton into a pair of gravitinos is given by~\cite{Endo:2006tf},
\begin{eqnarray}
   \Gamma(\phi \rightarrow 2\psi_{3/2}) \;\simeq\;
   \frac{|{\cal G}_\Phi^{\rm (eff)}|^2}{288\pi}  \frac{m_\phi^5}{m_{3/2}^2  
M_P^2},
   \label{eq:gamma_gravitino_GX}
\end{eqnarray}
with
\bea
|{\cal G}_\Phi^{\rm (eff)}|^2 
&\simeq&\left|\sqrt{3}\,K_{\phi \bar z} \frac{m_z^2}{m_\phi^2} \right|^2 
+ \left|3 (K_\phi - K_{\phi z {\bar z}})  \frac{m_{3/2}m_z^2}{m_\phi^3}\right|^2.
\eea
We have assumed that the diagonal elements of the kinetic terms are normalized as
\beq
K_{\phi {\bar \phi}} = K_{z {\bar z}} = 1,
\eeq
and that the kinetic mixing is small, $|K_{\phi {\bar z}}| \ll 1$.
%
Thus, we obtain
\beq
  \Gamma(\phi \rightarrow 2\psi_{3/2}) \;\simeq\; \frac{|K_{\phi {\bar z}}|^2}{96 \pi} \frac{m_\phi m_z^4}{m_{3/2}^2 M_P^2}+
   \frac{c'^2}{32\pi} \lrfp{m_z}{m_\phi}{4} \lrfp{\la \phi \ra}{M_P}{2}  \frac{m_\phi^3}{ M_P^2},
   \label{app1}
\eeq
where we have  defined
\beq
\la K_\phi - K_{\phi z {\bar z}} \ra \;\equiv  c' \la \phi^\dag \ra.
\eeq
In general, we expect $c' = {\cal O}(1)$ in the Planck unit. 
The first term in (\ref{app1}) is important only for light $m_z$ and heavy $m_\phi$, and so, we have focused on the
second term in the text.

\subsection{Decay into the scalar components of $z$}
Let us estimate the inflaton decay into $z$ and $z^\dag$. The decay into $zz$ is suppressed by the 
VEV of $z$. The effective interactions are obtained by expanding the kinetic term and the mass
term of $z$ as
\beq
{\cal L}\;=\; - K_{\phi z {\bar z}} \phi z \partial^2 z^\dag - e^G G^z G^{\bar z} 
\left(K_\phi K_{z \bar{z} z {\bar z}} + K_{z {\bar z} z {\bar z} \phi} \right) \phi z z^\dag + {\rm h.c.},
\eeq
where $G = K + \ln|W|^2$ and $|G_z| \simeq |G^z| \simeq \sqrt{3}$.
Note that the second terms is obtained by expanding the mass term for $z z^\dag$ with respect to $\phi$.
Using the equation of motion for $z$, the effective interactions can be written as
\beq
{\cal L} = - m_z^2 {\tilde c} \la \phi^\dag  \ra \phi z z^\dag,
\eeq
where we have used the fact that the mass of $z$ is given by
\beq
m_z^2 \;\simeq\; - e^G G^z G^{\bar z} K_{z {\bar z} z {\bar z}},
\eeq
and  we have defined
\beq
\la K_\phi-K_{\phi z {\bar z}} + \frac{K_{z {\bar z} z {\bar z} \phi}}{K_{z {\bar z} z {\bar z}}} \ra \;\equiv\; {\tilde c} \la \phi^\dag \ra.
\eeq
In general ${\tilde c} = {\cal O}(1)$.
The decay rate is thus given by
\beq
  \Gamma(\phi \rightarrow z z^\dag)  \;\simeq\; \frac{{\tilde c}^2}{16 \pi} \lrfp{\la \phi \ra}{M_P}{2} \lrfp{m_z}{m_\phi}{4} \frac{m_\phi^3}{M_P^2}
  \sqrt{1-\frac{4m_z^2}{m_\phi^2}}.
\eeq



\end{document}